\def \myplotone#1 {\centerline{#1}}

\def \maperture {3.5 \times 10^{14} h^{-1} M_{\sun}}
\def \virialmaperture {1.15 \times 10^{14} h^{-1} M_{\sun}}
\def \massfactor {3}
\def \masspergalaxy {8 \times 10^{12}h^{-1}M_{\sun}}
\def \approxOmegaval {2}
\def \Omegaval {1.9}
\def \precision {0.9\%}
\def \srms {0.43}
\def \ngalsused {2147}
\def \sigmarms	{0.022}
\def \sigmapeak	{0.11}
\def \nupeak	{5}
\def \betabright {0.27}
\def \betafaint {0.42}
\def \deltabeta {0.11}

\def \fullsamplebetaval {0.34}
\def \Sigmacritval {1.68}

\def \sigmabarval {6.0\%}
\def \sigmabarerror {1.3\%}
\def \dndz {8.2\times 10^7}
\def \netsone {0.37\%}
\def \netstwo {-0.38\%}
\def \rcore	{125 h^{-1} {\rm Mpc}}

\def \vectheta {{\vec \theta}}

\def \Psh {P^{\rm sh}}
\def \Psm {P^{\rm sm}}
\def \etal {{\it et al.\/}}
\def \Sigmacrit {\Sigma_{\rm crit}}
\def \Sigmarms {\Sigma_{\rm rms}}
\def \sigmabar {\overline\sigma}
\def \thetamax {\theta_{\rm max}}

\def \VIRIAL {Davis and Peebles, 1983}
\def \FLOWS  {Dekel, 1994}
\def \TVW	{Tyson, Valdes and Wenk, (1990)}
\def \MEA	{Miralda-Escude, 1991a}
\def \EMSS	{Gioia and Luppino, 1994}
\def \IMCAT	{Kaiser and Squires, 1994}

\def \MOULDETAL	{Mould \etal, 1994}
\def \KS	{Kaiser and Squires, 1993}
\def \LSSLENSREF	{(Blandford \etal, 1991; Miralda-Escude, 1991b; Kaiser, 1992)}

\documentstyle[aaspp]{article}
\begin{document}

\title{Dark Matter in ms1224 from Distortion of Background Galaxies}
\author{G.G.\ Fahlman\altaffilmark{1,2},
Nick Kaiser\altaffilmark{3},
Gordon Squires\altaffilmark{4,2}
and David Woods\altaffilmark{1,2}}
\altaffiltext{1}{Dept.\ of Geophysics and Astronomy, University of British
Columbia\\
2219 Main Mall, Vancouver, B.C. V6T 1Z4}
\altaffiltext{2}{Visiting Astronomer, Canada France Hawaii Telescope.
CFHT is operated by the National Research Council of Canada,
the Centre National de la Recherche Scientifique of France,
and the University of Hawaii}
\altaffiltext{3}{Canadian Institute for Advanced Research and\\
Canadian Institute for Theoretical Astrophysics, University of Toronto\\
60 St.\ George St., Toronto, Ontario, M5S 1A7}
\altaffiltext{4}{Physics Department, University of Toronto\\
60 St.\ George St., Toronto, Ontario, M5S 1A7}

\begin{abstract}
We explore the dark matter distribution in
ms1224.7+2007 using the gravitational distortion
of the images of faint background galaxies.
Projected mass image reconstruction reveals a highly significant concentration
coincident with the X-ray and optical location. The concentration is
seen repeatably in reconstructions from independent subsamples and
the azimuthally averaged tangential shear pattern is also
clearly seen in the data.
The projected mass within a $2.76'$ radius aperture is
$\simeq \maperture$.
This is $\simeq\massfactor$ times larger than that
predicted if mass traces light with $M/L = 275 h$ as derived from virial
analysis.
It is very hard to attribute the discrepancy to a statistical
fluctuation, and a further indication of a significant difference
between the mass and the light comes
from a second mass concentration which is again seen in independent
subsamples but which is not seen at all in the cluster light.
We find a mass per galaxy visible to $I = 22$ of $\simeq
\masspergalaxy$ which, if representative
of the universe, implies a density parameter $\Omega \sim \approxOmegaval$.
We find a null detection of any net shear
from large-scale structure with a precision of
\precision\ per component.  This is much smaller than the possible detection
in a recent comparable study, and the precision here is comparable to
to the minimum level of rms shear fluctuations implied by
observed large-scale structure.
\end{abstract}


\section{Introduction}
A puzzle in cosmology is the ``$\Omega$ discrepancy
problem'': virial analyses of clusters of galaxies give mass-to-light
ratios of typically a few hundred (\VIRIAL), implying $\Omega \simeq 0.1-0.2$,
while studies of supercluster scale dynamics (see e.g.\ \FLOWS\ for
a recent review) give larger values and seem compatible with $\Omega \simeq 1$.
One way to reconcile these results (and the theoretical
prediction that $\Omega = 1$) would be to argue that virial analysis
underestimates the cluster mass because the galaxies are more concentrated
than the dark mass.

This hypothesis is testable as we can measure the cluster mass directly
using the coherent shear of the images of distant background galaxies.
The observability of this effect was first demonstrated by
\TVW\ in A1689 and should be measurable, in principle, out to large distances
(\MEA; \KS).

Here we report a measurement of the image shear induced by the
cluster ms1224.7+2007.  This cluster was selected from the Einstein Medium
Sensitivity Survey (see \EMSS), has $L_x = 4.61
\times 10^{44}$ erg/sec ($h=0.5$) and lies at $z = 0.33$.  The cluster
has recently been extensively mapped in redshift space
by Carlberg, Yee and Ellingson (1994; hereafter CYE).
While the cluster is very bright in X-rays, it does not appear to
be particularly rich, shows no giant arcs and has a rather modest
velocity dispersion $\sigma \simeq 750$ km/s.  Nonetheless, as we will
show, we are able to recover a clear signal of weak lensing for this
cluster, and we draw some rather interesting conclusions regarding
the mass in this cluster and in the universe as a whole.
Bonnet \etal\ (1994) and Mellier \etal\  (1994) have recently
reported a detection of the shear near Q2345+007, and
several other groups (Tyson, 1994; Dahle \etal, 1994; Smail  \etal, 1994)
are currently pursuing similar studies.

\section{Data Analysis}

The data were acquired on the night of March 22/23, 1993 at the 3.6m CFHT.  The
detector
was the $2048 \times 2048$ Lick 2 chip at prime focus with a pixel scale
of $0.207''$.
We observed a grid
of four fields with corners overlapping the centre of the cluster.
Each field was observed in the $I$-band three times, with
slight positional offsets, for a total
integration time of 3600 seconds per field. The seeing for much of the night
was excellent (FWHM$\simeq 0.5''$).
At $z = 0.33$,
$1' = 0.174 h^{-1}$Mpc in physical units ($\Omega = 1$, $\Lambda = 0$) and the
observations cover a square approximately 2 Mpc/$h$ on a side
and centred on the cluster.

The 12 exposures allowed the construction of an accurate
median sky flat. The data were analysed with software
which will be described in detail elsewhere (Kaiser and Squires, 1994).
The analysis is a three step procedure. First we smooth each
exposure with a 2 pixel gaussian and locate all the peaks above
a low threshold.  Second, noise peaks, cosmic rays
etc are eliminated by requiring positional coincidence on overlapping
images.
Finally, we go back to
the unsmoothed images and analyse the pixels around the smoothed peak
location to determine shape and luminosity parameters.
To determine the luminosity and half light radius we use an
aperture of radius $\min(3r_p, 6\;{\rm pixels})$ where
$r_p$ is the `Petrosian radius' where $l(<r) / r$ peaks.
We analyse each exposure separately
and then average together the catalogue entries for the multiply
observed objects.  This avoids the problem of psf anisotropy introduced when
exposures are registered and added.

\begin{figure}  \label{fig:rlplot}
\myplotone{fig1.ps}
\caption{Size-magnitude distribution.  We find
$\simeq 5100$ objects in total
on about 140 square arcmin,
of which $\simeq 3200$ were brighter
than $I = 23.4$.
The box shows our subsample of $\simeq 2500$ faint galaxies.}
\end{figure}

Gravitational lensing will distort the surface brightness of background
galaxies according to
$f'(\theta_i) = f (\theta_i + \phi_{ij} \theta_j)$, where
the {\it image shear tensor} $\phi_{ij} $ is the
second derivative of the surface potential:
$\phi_{ij} \equiv \pi^{-1} \partial^2 \Phi/\partial \theta_i \partial \theta_j$
where $\nabla^2 \Phi = - 2\pi  \Sigma / \Sigmacrit$.
For an Einstein -- de Sitter universe, the effective critical
surface density is $\Sigmacrit = (4 \pi G a_l w_l \beta)^{-1}$, where
$\beta \equiv \langle \max(0,  1 - w_l / w_g) \rangle$
with subscripts $l, g$ denoting the lens and
background galaxies.
We define comoving distance as $w \equiv 1 - (1+z)^{-1/2}$ and
the scale factor is $a(w) = 6000 (1-w)^2 h^{-1}$Mpc.
To determine the shear coefficients
$s_\alpha \equiv
\{\phi_{11} - \phi_{22}, 2 \phi_{12} \}$
we use quadrupole moments (following \TVW).  Defining
$Q_{ij} = \int d^2 \theta \theta_i \theta_j f(\vectheta)$, and
image polarisation $e_\alpha \equiv
\{(Q_{11} - Q_{22}) / (Q_{11} + Q_{22}), 2 Q_{12} / (Q_{11} + Q_{22}) \}$,
it is easy to show that a small gravitational shear induces a
shift in the mean polarization $\langle e_\alpha \rangle = s_\alpha$,
so we can use $\hat s_\alpha = \sum e_\alpha  / N$
as an estimate of the shear with
statistical uncertainty
$\sqrt{\langle e^2 \rangle / 2 N}$ where $\langle e^2 \rangle$ is the rms
intrinsic polarisation and $N$ the number of objects.

The moments as defined above illustrate the idea, but are not practical
due to divergent sky noise.  To avoid this we
define
$Q_{ij} = \int d^2 \theta W(\theta) \theta_i \theta_j f(\vectheta)$.
The optimal weight function is a compromise between statistical
precision
and seeing.
We have used a gaussian window with scale length of 3 pixels which
was found, from
experiments with synthetic data (see below), to give
good results.
With $W(\theta) \ne 1$ we have
$\langle e_\alpha \rangle = \Psh_{\alpha\beta} s_\beta$ where
the {\it linear shear polarisability} $\Psh_{\alpha\beta}$
measures the response of the polarisation parameters $e_\alpha$
to a gravitational shear (\IMCAT).
We measure $\Psh_{\alpha\beta}$ for each object.
It is typically quite close to diagonal and we have used as our
estimator of the shear $\hat s_\alpha = \langle e_\alpha / \Psh \rangle$ where
$\Psh \equiv (\Psh_{11} + \Psh_{22}) / 2$.

Seeing will circularise the images. To calibrate this effect
and also to test the software we took samples from the actual
data, shrank these by a factor two, applied a known shear
and then convolved these images with a gaussian filter to mimic seeing
and added noise so that the faintest objects recovered were about as numerous
as
in the real data.  We are effectively modelling the faint galaxies as
scaled down replicas of somewhat brighter galaxies which are,
in reality, well resolved.
Analysing an ensemble of such images made without the seeing smoothing
we confirmed that the method does indeed recover the input shear
reliably.  With realistic seeing included we found that the output
was about 65\% of the input value.  In fact, the galaxies found from
the scaled images are slightly
smaller than the real galaxies, and
artificial simulations indicate that the appropriate value
for the real data is
$\simeq$70\% and we correct the
observed shear estimates appropriately.
To estimate the shear we simply average the shear values
for all galaxies with $|\hat s| < 1.4$, the cut here removing a small number of
typically very noisy images.

The shear we are trying to measure here is expected to
be only a few percent.  Anisotropies
arising from aberration or imperfect guiding are
clearly a worry. (Another potential problem, field distortion from
the telescope optics, is negligible at CFHT.)
We do find a significant psf anisotropy for the foreground stars, but we
have corrected for this using the property that a small
psf anisotropy induces a shift in polarisation
$\delta e_\alpha = \Psm_{\alpha\beta} p_\beta$, where
$p_\alpha$ is a measure of the psf anisotropy, and
where $\Psm_{\alpha\beta}$ is the {\it linear smear polarizability}
(see \IMCAT\ for details).

We used $e_\alpha / \Psm$ values from
a sample of about 400 moderately
bright stellar images to determine $p_\alpha$.
We found two significant effects.  The first was a
pixel position independent $p_\alpha$ which is different
in different exposures.  The second effect was a gradient
across the chip which appears to be constant from exposure to exposure.
Both effects were substantial: The gradient term gave
stellar anisotropy of about 10\% at the edge of the chip relative to
the centre in the best seeing, and this would generate a polarization
of a few percent in a typical small faint galaxy.
The $\chi^2$ statistics for the two stellar
polarisation components dropped by a factor $\simeq 2.7$ when these terms are
removed.
Including a quadratic term in the model gave a negligible further improvement
so we have used the constant plus gradient model.
We have tested the correction method using synthetic images,
and we believe that any residual shear from psf
anisotropy is well below 1\%.

\section{Results}

{}From the
subset of $\simeq2500$ faint, non-stellar images
we have removed objects with $|\hat s| > 1.4$ and also
objects which, though real, showed relative positional
offsets of more than one pixel in $x$ or $y$ (the parent sample
shown in figure 1 includes objects with offsets
$\le 2$) as these seemed to
be much noiser in polarization, reducing the number to \ngalsused.
The shear estimates $\hat s_\alpha = e_\alpha / \Psh$ are
shown in figure 2, which gives an idea of the kind of precision attainable
here.
The rms shear (per component) was
$\langle \hat s_i^2 \rangle^{1/2} \simeq \srms$.

\begin{figure}  \label{fig:ssplot}
\myplotone{fig2.ps}
\caption{Shear estimates $s_\alpha$ for the faint galaxy subsample.
This shows, to first order, the intrinsic distribution of
shear estimates.  The effect of the cluster lens would
not appear in this plot, but any shear which is coherent over
the whole field (as produced by  superclusters close to the line
of site) would produce an asymmetric shift of the distribution.
It is clear that any such shift is very small.}
\end{figure}

A reconstruction of the mass surface density is
shown in figure 3.
This was made using the method of Kaiser and Squires (1993, hereafter KS),
though correcting a minor error --- the correct result
is one half that given in their equation (2.2.1).
The reconstruction should not be trusted near the edge of the
image, but
in the inner regions we see a strong mass concentration whose location
coincides
very well with the smoothed optical centroid and lies a
little to the East of the X-ray location.
The rms noise in the estimator
is $\Sigmarms = \sqrt{\langle s_i^2 \rangle / 16 \pi \overline n \sigma^2}
\Sigmacrit$,
where $\overline n$ is the surface number density of galaxies and $\sigma$ is
the
gaussian smoothing radius. For the smoothing used here we find $\Sigmarms
= \sigmarms \Sigma_{\rm crit}$.  The peak value is
$\Sigma \simeq \sigmapeak \Sigma_{\rm crit}$, a \nupeak-sigma upward excursion.
In figure 3 we have added a correction for finite data as described
below. This raises the peak height to $\simeq 0.13\Sigmacrit$,
but the statistical significance is essentially unaltered.
No correction has been made for contamination of our sample
by cluster members which will have diluted the signal in the central region.

\begin{figure}  \label{fig:massplot}
\myplotone{fig3.ps}
\caption{
The upper left panel shows a contour plot of the
surface density $\sigma(\vec \theta) \equiv \Sigma(\vec \theta) /
\Sigmacrit$
derived from the background galaxy shear estimates for our
full sample.
The circles show the aperture and control annulus for our $\sigmabar$ estimate.
The panels on the right are derived from independent subsamples.
A strong central mass concentration is clearly seen, and
the northerly extension or sub-clump is also seen repeatably.
The lower left panel shows the surface density
predicted from the CYE redshift survey data
assuming M/L = 275$h$ and with $\Sigmacrit$ as
appropriate for the full sample.  The dotted line shows the extent
of their data. All images have been
smoothed with a $40''$ gaussian filter.  The angular coordinates are
in arcminutes relative to the giant elliptical galaxy.
}
\end{figure}

The mass map in figure 3 is made by convolving the shear estimates with
a certain kernel.  We can see the signal in the data more directly
in figure 4a where we have plotted the mean tangential shear
$s_T = - \langle s_1 \cos 2\phi + s_2 \sin 2\phi\rangle$
where $\phi$ is the azimuthal angle of the background galaxy
position relative to the cluster centre (which we have taken to be the
peak of the mass map shown in figure 3), as a function of radius.
The signal is clearly seen at radii $\simeq 1'-5'$ (any contamination from
cluster members is very small at these radii).

\begin{figure}  \label{fig:stplot}
\myplotone{fig4.ps}
\caption{The upper panel shows the mean tangential shear
(corrected
for seeing) as a function of radius
relative to the peak of the mass distribution seen in figure 3.
The lower panel shows $\zeta(r)$, which provides a lower
bound on the mean surface density interior to $r$ (see text).
The dashed lines are the predictions for isothermal spheres with
velocity dispersions 700 and 1000 km/s.
}
\end{figure}

We now want to estimate the mass enclosed within a radius of
$2.76'$ (800 pixels) which seems to enclose the significant mass concentration.
A fundamental problem of weak lensing analysis is an ambiguity
in the base level of the surface density.  With finite data
the KS estimator underestimates the true surface density.  The correction
consists of two terms; the first is half the mean tangential shear at the
boundary. From figure 4 we can estimate this to be about 2\% and we have
applied this correction
in figure 3. The second term is the mean surface density at the boundary
which is not easily measured.  A better way to handle this problem is
to modify the estimator.  The KS estimator is a convolution of the shear
estimates with the shear pattern from a point mass.  If we truncate the
kernel at inner and outer radii $r_1,r_2$ and divide
by $(1 -  r_1^2/r_2^2)$  the result is a convolution of the
surface density with a compensated top-hat; i.e.\ it measures the
mean surface density within a disk of radius $r_1$ minus that in
a control annulus from $r_1$ to $r_2$.  Since the latter is positive,
this gives a lower bound (barring noise) on $\sigmabar(r_1)$,
the mean surface density in units of $\Sigmacrit$ interior to $r_1$.
A nice feature of this statistic is that is only uses data at
$r > r_1$, largely avoiding the problem of dilution by cluster
members.
Evaluating this at the peak of the mass map, we have
$\sigmabar(\theta) > \zeta(\theta) \equiv
(1 - \theta^2 / \thetamax^2)^{-1} \int_\theta^{\thetamax}
d \ln \theta\; s_T$, which is plotted in figure 4b.
For $\theta = 2.76'$ this gives $\sigmabar = \sigmabarval \pm
\sigmabarerror$.

To convert $\sigmabar$ to an estimate of the physical mass
we need to estimate $\Sigmacrit$. This depends, through $\beta$, on the
redshift distribution of our galaxies which is still imperfectly
known; in fact the uncertainty in $\beta$ dominates our error budget.
For the brighter half of our galaxies
we can estimate $\beta$ directly from the complete redshift
samples of Lilly, 1993 and Tresse, \etal, 1993 and we find,
for $I = 20-22$, $\beta \simeq \betabright$, with statistical
error $\delta\beta \simeq \deltabeta$.
To get the appropriate
value for the fainter half we need to extrapolate.  Splitting the
$I = 20-22$ surveys in two, we find an increase in $\beta$ of about
30\% per magnitude
suggesting $\beta \simeq \betafaint$ for
our faintest magnitude slice (though the data are consistent with
no increase in depth at the 1-sigma level),
and we find, for the whole
sample, $\beta = \fullsamplebetaval$ and
therefore $\Sigmacrit = \Sigmacritval h$g cm$^{-2}$.

With this $\beta$-value,
the total projected mass within our $2.76'$ aperture is
$\simeq\maperture$. Let us now compare this with the light.
The lower panel in figure 3 shows the surface
density predicted by CYE assuming light traces mass with
$M/L = 275h$ as inferred from virial analysis.
While the positions of the peaks agree nicely, the shear-derived
mass peak is higher and perhaps more extended than the light.
According to the virial estimates the mass within our $2.76'$
aperture is $\virialmaperture$, which is a factor $\massfactor$
lower than our estimate.

This is a large discrepancy, so it is worthwhile pausing to consider
possible biases and uncertainties in our method.  We are aware of a number
of biases, but all of these will have tended to
cause us to underestimate the mass.  First, as mentioned, our
estimator tends to underestimate $\sigmabar$ if there
is mass in the cluster which extends beyond the aperture.
A further bias is introduced if, as done in the light-map here, we
include material seen in projection in our aperture which is
at a different distance without allowing for material in the control
annulus at similar distance --- this is true whether or not the material
is physically associated with the cluster --- and removing
this increases the discrepancy by about 10\%.
Finally, there is some reason to suspect that our extrapolation
of $\beta$ for the fainter galaxies is an overestimate as we
can infer the relative $\beta$ values from the strength of the
lensing signal and this exercise suggests
a further increase of about $10-20\%$ in the mass estimate.
Thus our estimate of the discrepancy
is in many respects a cautious one, and the real discrepancy could
well be even larger.
The discrepancy is hard to attribute to statistical fluctuation
in the shear estimates; it is about
3 times our statistical error.  There is also uncertainty in $\beta$, but to
account
for the discrepancy with this alone would require $\beta \simeq 0.76$.
This is about 4-sigma removed from our estimate, and
corresponds to a typical redshift $z \simeq 4$ for the sample as a whole,
and only slightly smaller values for our brighter galaxies.
This is clearly incompatible with the data of Lilly and Tresse \etal.
If we use the brighter subsample alone, and thus avoid the need to
extrapolate, we find $M \simeq 4.5 \pm 1.3 \times 10^{14} h^{-1} M_{\sun}$
which is somewhat noisier, but still discrepant at the 2.5-sigma level
with the virial estimate.

Our large total mass implies an enormous $M/L \sim 800$.  This, coupled with
the well established increase in comoving luminosity density of the universe
at these redshifts, implies a very large value for the density parameter if
this is estimated in the usual manner (i.e.\ by dividing the cluster $M/L$ by
the $M/L$ for a critical density universe).
As the luminosity density of the universe is very noisy with the current
samples, it is preferable to calculate a mass-per-galaxy and compare
this with the mass-per-galaxy for a $\Omega = 1$ universe.
Comparing the galaxy counts in our aperture with that in the surrounding
annulus --- note that this avoids the bias problem --- we
see
an excess of $N_c \simeq 45 \pm 16$ cluster galaxies to $I = 22$
within our aperture.
This gives a mass $\simeq \masspergalaxy$ per galaxy.
Assuming that the number of galaxies per unit mass in the
cluster is representative of that of the universe, we obtain an
expression for the density parameter:
\begin{equation}
\Omega = {
\overline \sigma d\Omega (1+z_l)^{1/2} dn / dz_l
\over 3 N_c w_l \beta
} \simeq \Omegaval
\end{equation}
$\delta \Omega$ is the solid angle of the aperture, and
$dn/dz$ is the differential
counts per unit redshift per steradian in the appropriate magnitude range,
which we estimate from Lilly (1993) and Tresse \etal\ to be $\simeq \dndz$
at $z = 0.33$.

These data can also be used to place limits on
shear from large-scale structures \LSSLENSREF.
A similar study to that performed here, though not for a cluster
field (\MOULDETAL), gave an apparent detection of net shear
$|s| \simeq 5$\% when corrected for seeing,
though the authors felt unable to reject the possibility
that their detection may be a residual artefact.
Here we obtain a net shear
$\hat s_\alpha =
\{ \netsone\pm \precision, \netstwo\pm \precision \}$
which is a null detection
of much greater sensitivity.

The \MOULDETAL\ exposures were deeper than ours, and
the rms image shear is predicted to increase as $s_{\rm rms}
\propto \langle w^3\rangle^{1/2}$.
They adopt a median redshift $\overline z \simeq 0.9$ ($\overline w = 0.27$).
{}From the redshift survey data we obtain $<w^3>^{1/3} \simeq 0.23$
which is a little lower and would suggest
a $\sim25$\% decrease in cosmological signal.
Another way to estimate the relative comoving depths is to compare
the number density of galaxies above the respective magnitude
limits; theirs was about 2.7 times higher than here.
At brighter magnitudes, $\overline w$ is
a rather weak function of number density:
$\overline w \propto n^{0.2}$, so if this trend continues the prediction
for their survey should be about 30\% larger than for ours for any given
model, consistent with the estimate obtained using their estimated
median redshift.

The present sensitivity $\langle s^2 \rangle^{1/2} \simeq 1.3$\%
is roughly a factor 4 lower than
the Mould \etal\ result, and so, taking their possible detection as a measure
of their
sensitivity, our result represents an increase of about a factor 3
in sensitivity to cosmological signal.

The sensitivity we have reached is comparable to the minimum rms shear
compatible with the large-scale mass fluctuations implied by
bulk flows (Kaiser, 1991).  High normalisation CDM models with
$\Omega = 1$, $\sigma_8 = 1$ give predictions of a few percent shear,
which should be easy to measure or exclude with further
observations of this kind.
It is difficult to draw strong cosmological conclusions on the basis of
the two components of the net shear from a single field --- and in any case,
one would want to avoid fields such as this one containing a cluster --- but
the results
here show that with more fields of similar precision a direct measurement
of large-scale mass fluctuations should be feasible.

\section{Summary}

Shape analysis of $\sim$2000 galaxies behind ms1224 clearly reveals the
gravitational image shear from the cluster and has allowed
us to reconstruct the projected mass, whose main peak coincides
to within about one arcminute with the optical and X-ray centroids.
Our most surprising result is that the projected mass we infer is much larger
that implied assuming mass traces light with the virial mass-to-light estimate
$M/L = 275h$.  We have argued that it is hard to account for the discrepancy
by statistical fluctuation as it would seem to require both that
we were the victim
of a large positive fluctuation and that the redshifts in current samples
at $I \sim 20-22$ are unrepresentatively low.  This improbable
possibility can be eliminated or confirmed with future observations.

The discrepancy is large, and appears at a surprisingly small radius.
Our estimator effectively measures the mean projected surface density within
our aperture (of
about 0.5$h^{-1}$Mpc radius) relative to an average value at about twice this
radius.
Adding a halo in a spherical manner at larger radii --- which would obviously
not conflict with the virial analysis --- does not influence our statistic,
and adding shells of mass in the control region will actually
register as negative mass due to limb brightening.

It is difficult to say whether the light is more concentrated than the
mass within our annulus. The shear in the central region is rather weak,
and considerably weaker than a singular isothermal sphere model matched
to the shear at large radius for instance.  This would seem to require
a core radius of $\simeq\rcore$, but we caution that the shear in the
inner region is diluted by cluster members.

A further piece of information which argues in favour of a
significant discrepancy
between mass and light is the second peak or extension in the mass
reconstruction. This feature appears in both bright and faint subsample
reconstructions but is not seen at all in the cluster light.
With much deeper observations it should be possible,
in principle, to reduce the statistical error
in the mass-map by potentially a factor 3 or so.

We find a mass-per-galaxy of $\simeq\masspergalaxy$.
If this is representative of the universe at large --- which of course
it need not be ---
then $\Omega \simeq \approxOmegaval$.  The precision of this estimate, like
that for the
cluster mass, is currently limited by the paucity of faint galaxy redshifts and
will improve considerably as more redshift information
becomes available.

We have paid particular attention to correcting for the effect of
anisotropy of the psf.  This, and the generally superior
image quality at CFHT, has allowed a sensitivity
to shear from large-scale structure which is much greater than
previously obtained.  While the field here is not ideal for
that purpose, we have shown that directly measuring mass
fluctuations on scales of order tens of Mpc in this way is indeed feasible.

\acknowledgements{We thank Ray Carlberg, Erica Ellingson and Howard Yee for
access to results prior to publication.  We acknowledge many useful
conversations with
Simon Lilly, Jens Villumsen, Roger Blandford and Tony Tyson.
This research was supported in part by operating grants from the
National Science and Engineering Research Council of Canada.
We thank the director of CFHT for the allocation of telescope time.
}


\begin{references}
\reference Blandford, R.D., Saust, A.B., Brainerd, T.G. and Villumsen, J.V.,
1991. \mnras, {\bf 251}, 600
\reference Bonnet, H., Fort, B., Kneib, J.-P.,  Mellier, Y. and Soucail, G.
1994.
\aap, {\bf 280}, L7
\reference Carlberg, R., Yee, H., and Ellingson, E.,  1994. in preparation
\reference Dahle, H., Maddox, S.J. and  Lilje, P.B., 1994. in preparation
\reference Davis, M. and Peebles, P.J.E, 1983. \araa, {\bf 21}, 109
\reference Dekel, A., 1994. \araa, {\bf 32}
\reference Gioia, I. and Luppino, G., preprint
\reference Kaiser, N., 1992. \apj, {\bf 388}, 272
\reference Kaiser, N. and Squires, G., 1993. \apj, {\bf 404}, 441
\reference Kaiser, N. and Squires, G., 1994. in preparation
\reference Lilly, S.J., 1993. \apj, {\bf 411}, 501
\reference Mellier, Y., Fort, B., Bonnet, H., and  Kneib, J.-P. 1994.
to appear in ``Cosmological Aspects of X-Ray Clusters of Galaxies" NATO
Advanced
Study Institute, W. Seitter et al eds.
\reference Miralda-Escude, J., 1991a. \apj, {\bf 370}, 1
\reference Miralda-Escude, J., 1991b. \apj, {\bf 380}, 1
\reference Mould, J., Blandford, R., Villumsen, J., Brainerd, T.,
Smail, I., Small, T. and Kells, W., 1994. preprint
\reference Smail, I., Ellis, R.S., Fitchett, M.J.\ \& Edge, A.C., 1994. in
preparation
\reference Tresse, L., Hammer, F., le Fevre, O., and Proust, D., 1993
\aap, {bf 277}, 53
\reference Tyson, J., Valdes, F. and Wenk, R., 1990, \apjl, {\bf 349}, L19
\reference Tyson, J., 1994. in proceeding of Les Houches summer school
\end{references}
\end{document}